\documentclass[reprint, amsmath, amssymb, aps, superscriptaddress, prb]{revtex4-2}
\usepackage{amsmath}
\usepackage{amssymb}
\usepackage{hyperref}

\usepackage{graphicx}
\usepackage[normalem]{ulem} 
\usepackage{lipsum}

\usepackage{xcolor}

\begin{document}

\begin{abstract}
We study the quantum Hall effect (QHE) in the three-dimensional topological insulator HgTe, which features topological Dirac-type surface states in a bulk gap opened by strain. Despite the co-existence of multiple carrier subsystems, the system exhibits perfectly quantized Hall plateaus at high magnetic fields. Here we study the system using three different experimental techniques: Transport experiments, capacitance measurements including the quantum capacitance, and current distribution measurements using electrostatically sensitive scanning probe microscopy. Our key finding is that at sufficiently high magnetic fields, the different electronic subsystems merge into one, and the current in a quantum Hall plateau is distributed across the entire width of the Hall bar device.
\end{abstract}

\title{Quantum Hall effect and current distribution in the 3D topological insulator HgTe}

\author{S.~Hartl}
\affiliation{Experimental and Applied Physics, University of Regensburg, D-93040 Regensburg, Germany}

\author{L.~Freund}
\affiliation{Max Planck Institute for Solid State Research, Heisenbergstrasse 1, D-70569 Stuttgart, Germany}

\author{M.~K{\"u}hn}
\affiliation{Max Planck Institute for Solid State Research, Heisenbergstrasse 1, D-70569 Stuttgart, Germany}

\author{J.~Ziegler}
\affiliation{Experimental and Applied Physics, University of Regensburg, D-93040 Regensburg, Germany}

\author{E.~Richter}
\affiliation{Experimental and Applied Physics, University of Regensburg, D-93040 Regensburg, Germany}

\author{W.~Himmler}
\affiliation{Experimental and Applied Physics, University of Regensburg, D-93040 Regensburg, Germany}

\author{J.~B{\"a}renf{\"a}nger}
\affiliation{Experimental and Applied Physics, University of Regensburg, D-93040 Regensburg, Germany}

\author{D.\,A.~Kozlov}
\affiliation{Experimental and Applied Physics, University of Regensburg, D-93040 Regensburg, Germany}
\affiliation{Rzhanov Institute of Semiconductor Physics, 630090 Novosibirsk,
	Russia}

\author{N.\,N.~Mikhailov}
\affiliation{Rzhanov Institute of Semiconductor Physics, 630090 Novosibirsk,
	Russia}

\author{J.~Weis}
\affiliation{Max Planck Institute for Solid State Research, Heisenbergstrasse 1, D-70569 Stuttgart, Germany}

\author{D.~Weiss}
\affiliation{Experimental and Applied Physics, University of Regensburg, D-93040 Regensburg, Germany}

\maketitle

The Quantum Hall effect (QHE) stands as a defining characteristic of two-dimensional electron systems (2DES), and its manifestations have been documented in a multitude of diverse systems, like silicon MOSFETs \cite{Klitzing1980}, semiconductor heterostructures \cite{Guldner1982,Klitzing1983} and quantum wells \cite{Altarelli1987, Hopkins1991} or graphene \cite{Novoselov2005, Novoselov2006, Zhang2005}. This groundbreaking effect plays a pivotal role in precision metrology, leading to the redefinition of SI units in 2019 \cite{Klitzing2019}. The QHE is the first experimental manifestation of the role of topology in condensed matter science \cite{Thouless1982}. 
A recent intriguing observation of the QHE has occurred in three-dimensional topological insulators (3D TIs), such as strained HgTe layers \cite{Brune2011, Kozlov2014}, Bi-based materials \cite{Xu2014, Yoshimi2015}, or $\beta$-Ag$_2$Te \cite{Leng2023}. In contrast to conventional 2DES the two-dimensional electron system in a 3D TI consists of the Dirac surface states that encompass the bulk of the material \cite{Ando2013}  and has a different geometric topology. Thus, different charge carrier species on the different surfaces, and bulk electrons or holes, depending on the Fermi level position, contribute to the electrical transport \cite{Ziegler2020}.

Here we study the QHE in strained 3D HgTe, which is a strong 3D topological insulator \cite{Fu2007} and exhibits high charge carrier mobilities of several $10^5$\,cm$^2$/V$\cdot$s \cite{Kozlov2014}. On the one hand, we study how it can be that a system with several charge carrier systems has a pronounced quantization of the Hall resistance $R_{xy}$ \cite{Lee2009} and, on the other hand, how the current is distributed in the quantum Hall regime. The latter is motivated by the fact that theories stress the role of edge states at the side facets oriented parallel to the magnetic field $B$ or at the side edges of a 3D TI (see, e.g., references \cite{Lee2009, Vafek2011, Zhang2011, Sitte2012}). 

\begin{figure}
	\includegraphics[width=1\columnwidth]{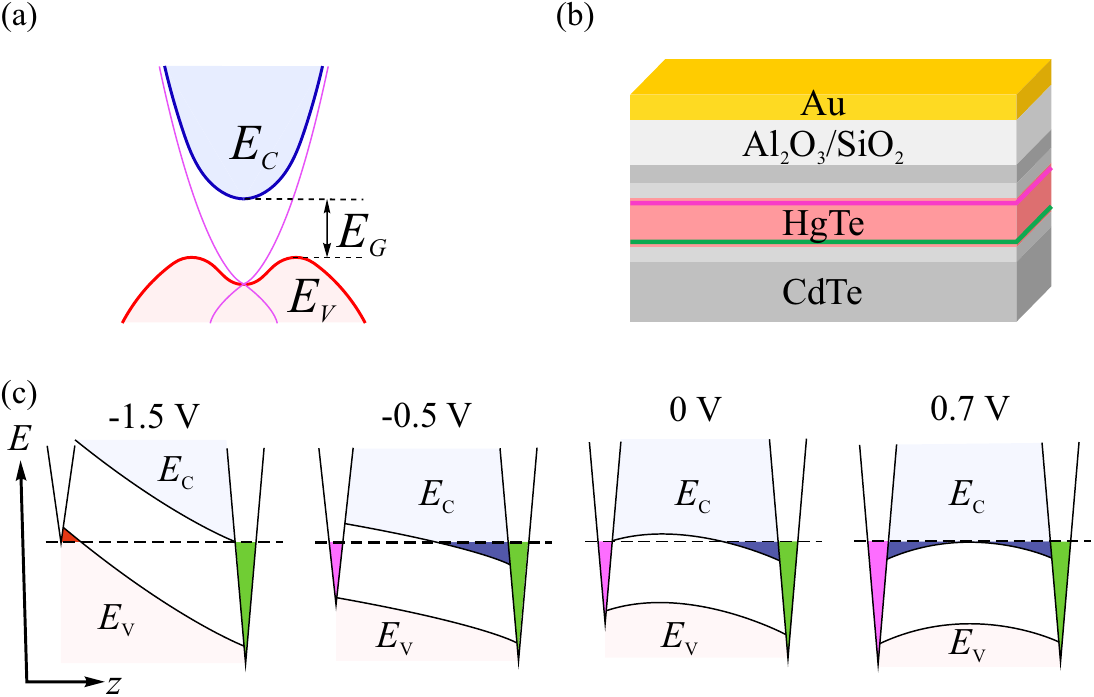}
	\caption{
		(a) Simplified electronic band structure of tensile-strained HgTe around the $\Gamma$ point as a function of an in-plane wave vector. 
		The conduction band with bottom $E_{\rm C}$ is shown in blue, the valence band with top $E_{\rm V}$ in red and the topological surface states in magenta. The bulk gap $E_{\rm G}$, opened by strain, is of order 15\,meV. 
		(b) Heterostructure layers grown on (013)-GaAs and 5\,$\mu$m CdTe containing the 80\,nm thick strained HgTe film. 
		Cladding layers of 20\,nm thick Cd$_{0.68}$Hg$_{0.32}$Te below and 20\,nm thick Cd$_{0.60}$Hg$_{0.40}$Te above HgTe are doped with $10^{17}$\,cm$^{-3}$ In distributed in a 10\,nm layer. 
		The next layers in the growth direction are 40\,nm CdTe, 30\,nm SiO$_2$ and 80\,nm Al$_2$O$_3$, a 5\,nm Ti adhesion layer, and 70\,nm Au, which serves as the top gate. The topological surface states in HgTe are shown by the magenta (top) and green (bottom) lines. 
		The top gate allows tuning the carrier density and Fermi level by applying a DC voltage $V_g$. 
		(c) Schematic band bending through the HgTe layer for different gate voltages $V_g$. The top surface is on the left side, respectively, and the cones sketch its density of states. The dashed line indicates the Fermi level. Filled top surface states are shown in magenta, filled bottom surface states in green, filled bulk hole states in red, and filled bulk electron states in blue. Due to the doping on both sides, the conduction and valence bands are bent towards the surface.
}
	\label{Fig1_p1}
\end{figure}

The band structure of strained 3D HgTe is sketched in Fig.~\ref{Fig1_p1}(a) and shows the conduction band, the valence band, and the spin-resolved, gapless topological surface states. In strained HgTe, the Dirac point is buried in the valence band \cite{Brune2011, Crauste2013, Wu2014}. The molecular beam epitaxy grown layer structure containing the 80\,nm thick HgTe layer is shown in Fig.~\ref{Fig1_p1}(b). Since the lattice constants of HgTe and the underlying CdTe layer are different, the typically semi-metallic HgTe is tensile-strained by about 0.3\%, opening the gap between the valence and the conduction bands \cite{Brune2011, Crauste2013, Wu2014}. The topological surface states, which form a two-dimensional electron system wrapped around the bulk states, are indicated by the magenta (top surface) and green line (bottom surface). Not shown, they also appear at the etched mesa's side facets, directly connecting the top and bottom layers. A gold gate on top of the structure allows tuning the charge carrier densities.

Due to the three-dimensional nature of the system, different groups of charge carriers (subsystems) contribute to transport and react differently to the applied gate voltage. Fig.~\ref{Fig1_p1}(c) schematically shows the change in the cross-sectional band diagram as the gate voltage is varied. The cones on the left correspond to Dirac surface states on the top surface, i.e. near the gate electrode. Those on the right represent the Dirac electrons on the backside. For sufficiently negative $V_g$ (here -1.5\,V) the top surface Dirac states are depleted, bulk holes (in red) appear, and the bottom surface states are occupied up to the Fermi level. Corresponding data are shown in Fig.~\ref{FigA1} of the Appendix. Sweeping $V_g$ to positive values increases the top surface carrier density $n_{\rm top}$, and the bulk electrons (blue), while the backside Dirac electron density (green) remains nearly constant. This is due to the screening of the electric field of the gate electrodes by the charge carriers in between.

\begin{figure}
	\centering
	\includegraphics[width=1\columnwidth]{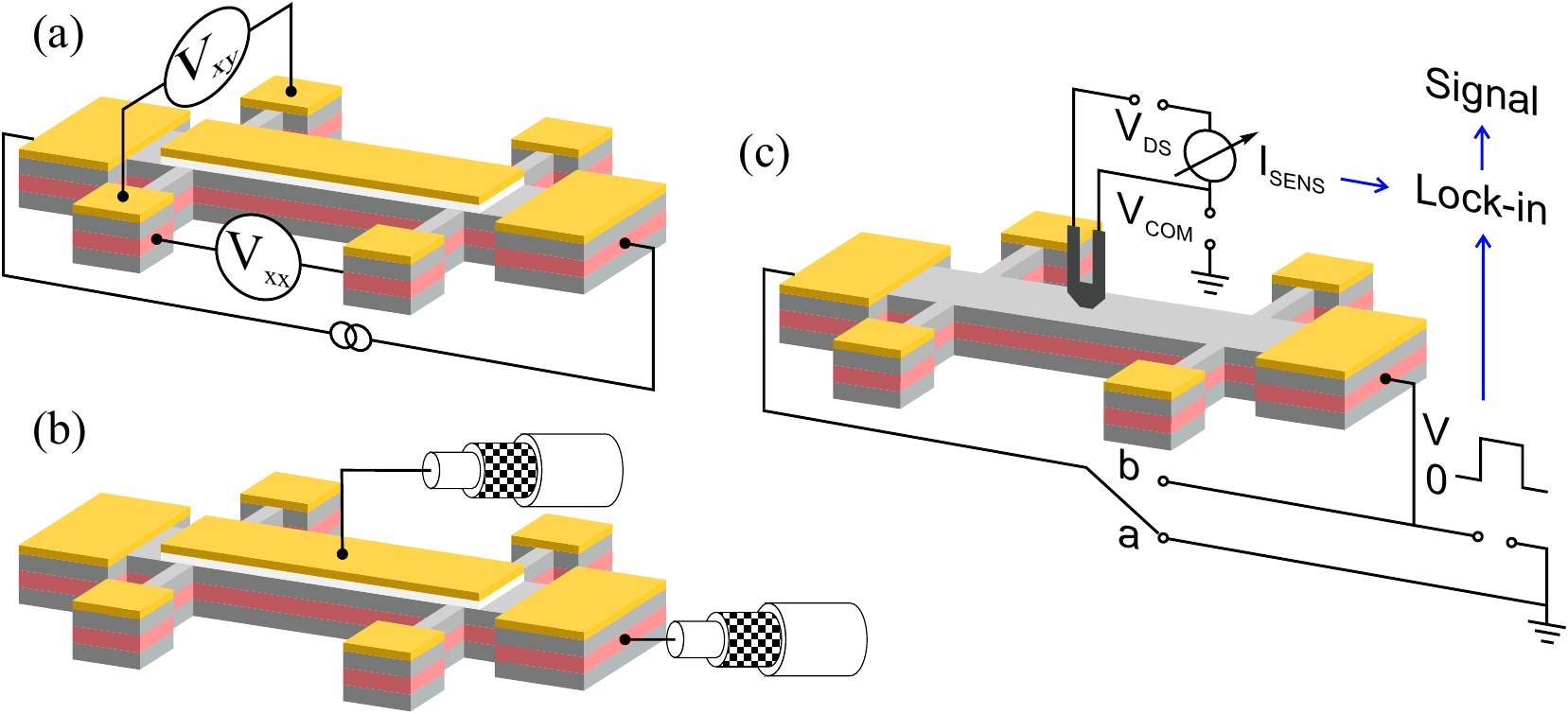}
	\caption{Three complementary techniques used to study QHE. 
		(a) Schematic gated Hall bar for measuring longitudinal and Hall resistance in 4-point geometry using standard lock-in techniques. For the transport experiments, we used alloyed indium contacts (not shown). 
		(b) The capacitance measuremnts were performed by applying an AC voltage $V_{\rm AC}$ on top of the DC gate voltage $V_g$ between the top gate and HgTe, indicated by the schematic coax cables. The resulting AC current is measured phase-sensitively using lock-in techniques. 
		(c) On an ungated Hall bar, electrostatically sensitive scanning probe microscopy (SPM) was applied: Scanning over the cross-section of the Hall bar twice while switching between two voltage excitation schemes, a and b, allows for a detailed mapping of the Hall potential profile and reflects the spatial non-equilibrium current distribution in the quantum Hall regime. The voltage $V_{\rm COM}$ is used to compensate for a workfunction difference, i.e., a local electrostatic depletion of the charge carriers within the HgTe layer by the probe tip is avoided. For contacting, the CdTe was partially removed before the Au contacts were deposited. 
	}
	\label{Fig1_p2}
\end{figure}

To study the QHE in this complex layered system, we used three complementary techniques, sketched in the panels of Fig.~\ref{Fig1_p2}: The electrical current in a 3D TI is carried by all charge carrier subsystems in the HgTe layer, and hence quantum oscillations and QHE are affected by all participating charge carriers. A schematic diagram of the transport measurements setup, using a Hall bar geometry of width 50\,$\mu$m and a contact spacing of 100\,$\mu$m, is shown in Fig.~\ref{Fig1_p2}(a). The respective measurements were done at 57\,mK.
Secondly, by driving an AC voltage $V_{\rm AC}$ superposed to a DC voltage $V_g$ between the top gate and the HgTe layer (Fig.~\ref{Fig1_p2}(b)) and measuring the displacement current phase-sensitively at sufficiently low frequencies with a lock-in amplifier, it is possible to extract the capacitance $C$. This $C$ includes, in addition to the geometric capacitance $C_{\rm geo}$, the quantum capacitance $e^2 D$, which is proportional to the density of states (DOS) $D$ at the Fermi level of the conductor facing the metallic top gate. 
Here, $e$ is the elementary charge. In its simplest form the relation reads: $C^{-1} =C_{\rm geo}^{-1}+(e^2 D)^{-1}$ \cite{Smith1985, Mosser1986}. For a 2DES with a perpendicularly applied magnetic field, a half-filled Landau level at the Fermi level leads to a maximum in $D$ and therefore in the capacitance. 
Recent experiments on 3D HgTe TIs have shown that the capacitance probes at small and intermediate magnetic fields the DOS of the top surface of the HgTe layer only \cite{Ziegler2020, Kozlov2016}. Thus, the periodicity of the quantum oscillations seen in the capacitance in not-too-high magnetic fields reflects exclusively the carrier density of the top surface. The capacitance measurements presented here under different magnetic fields and DC voltage bias $V_g$ between the top gate and the HgTe layer were performed at a temperature of 5\,K. Finally, to get information about the current distribution in our 3D TI under quantum Hall conditions we used a scanning field effect transistor, sketched in Fig.~\ref{Fig1_p2}(c), which probes the Hall potential profile within the cross-section of an \textit{ungated} 20\,$\mu$m wide Hall bar at 40\,mK. From the gradients in the Hall potential profile across the Hall bar one can derive the non-equilibrium current distribution in the quantum Hall regime as is discussed in more detail below. 

\begin{figure}[t]
	\centering
	\includegraphics[width=1\columnwidth]{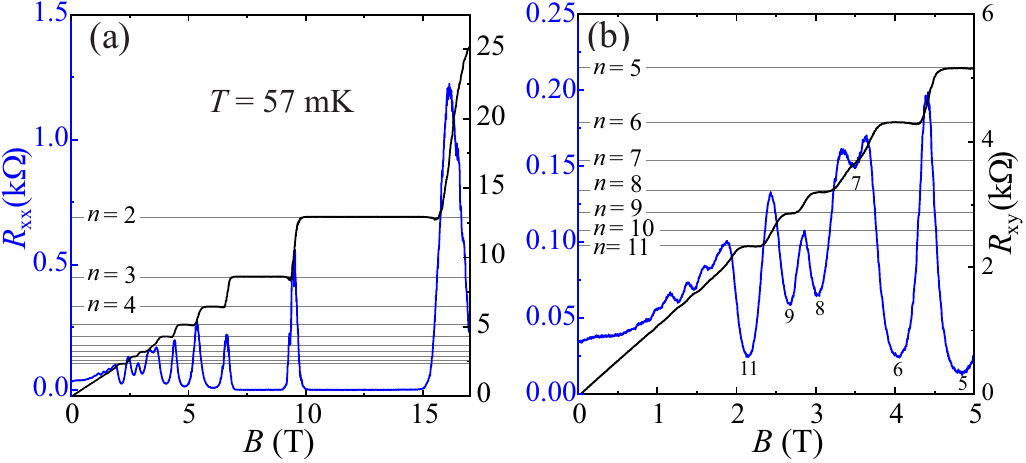}
	\caption{Hall resistance $R_{\rm xy}$ (black) and the square magnetoresistance $R_{\rm xx}$ measured at 57\,mK and $V_g= 0$ using an AC current of 10\,nA. (a) At high magnetic fields $R_{\rm xy}$ shows well quantized plateaus and zero resistance states in $R_{\rm xx}$. (b) Measurements at lower $B$ show a more complicated behavior, e.g., some of the quantum Hall plateaus and SdH minima are only faintly visible ($\nu  = 7$) or completely missing ($\nu = 10$), while others are still very pronounced. This reflects the coexistence of multiple charge carrier subsystems. The horizontal lines in both panels correspond to the expected position of plateau number $n$.
	}
	\label{Fig2}
\end{figure}

An example of transport data on our gated Hall bar, showing the QHE and the corresponding Shubnikov-de Haas (SdH) oscillations is shown in Figs.~\ref{Fig2}(a) and \ref{Fig2}(b), which differ in the magnetic field scale. At high magnetic fields in Fig.~\ref{Fig2}(a), $R_{\rm xy}$ shows well-quantized quantum Hall plateaus and $R_{\rm xx}$ pronounced zero-resistance states with $R_{\rm xx}$ spikes in between. The existence of well-quantized quantum Hall plateaus indicates that a two-dimensional charge carrier system dominates the transport in that respective parameter regime. 
From the plateau values at high magnetic fields, we can extract the Landau level filling factors $\nu = 2$ and $\nu = 3$. The $R_{\rm xx}$ spikes mark DOS maxima at the Fermi level which are expected for a half-filled Landau level (LL). At lower magnetic fields, better seen in Fig.~\ref{Fig2}(b), the SdH oscillations and associated $R_{\rm xy}$ plateaus show some irregularities: features for $\nu = 10$ are missing in both $R_{\rm xy}$ and $R_{\rm xx}$, and the SdH minimum for $\nu = 7$ and the corresponding Hall plateau is only weakly pronounced. 

\begin{figure}[t]
	\centering
	\includegraphics[width=0.8\columnwidth]{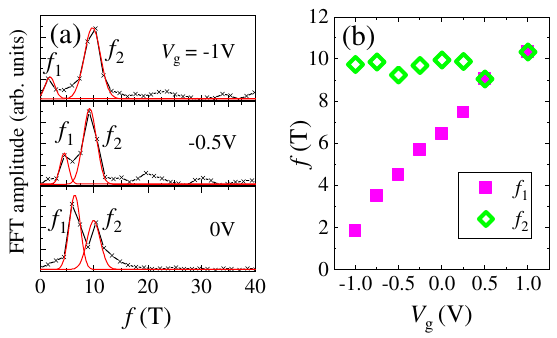}
	\caption{(a) Fast Fourier transform (black) of the low-field SdH oscillations from 0.75\,T to 1.5\,T at three gate voltages, which can be decomposed into two characteristic frequencies (red). (b) While $f_1$ increases linearly with $V_g$, $f_2$ remains constant.
	}
	\label{Fig3}
\end{figure}

We expect this irregular behavior because, at least at lower magnetic fields, there are three subsystems – occupied top/bottom surface states and bulk electrons or holes – with different evolution of Landau levels with a magnetic field (‘Landau fans’), and this coexistence should leave its mark on the transport data. By fast Fourier transforming (FFT) the low-field SdH oscillations in the voltage regime between -1\,V and 0\,V, where the Fermi level is above the valence band edge, plotted over $1/B$ \cite{Mendez1985, Savchenko2023}, we can obtain more information about the densities of the dominant charge carrier systems there. Fig.~\ref{Fig3}(a) shows the FFT amplitudes for three different gate voltages, with two peaks in the spectrum, as expected for the Fermi level in the gap. One of these frequencies (the inverse of the periodicity in $1/B$), $f_1$, increases linearly (Fig.~\ref{Fig3}(b)) with the gate voltage. The corresponding carrier density, given by $e f_1/h$ corresponds to the density of a carrier system directly facing the top gate – the Dirac electrons at the top surface. The other frequency, $f_2$, remains constant, and is hence ascribed to a carrier system which is screened from changes in $V_g$ – presumably the Dirac electrons at the bottom surface and/or electrons in the bulk facing the bottom surface (see band diagram in Fig.~\ref{Fig1_p1}(c)). The corresponding data points are shown in Fig.~\ref{Fig3}(b) and Fig.~\ref{FigA1} in the Appendix. For $V_g<0.5$\,V, $f_1$  is smaller than $f_2$, i.e., the surface electrons have less density than the bottom electrons. Somewhere between 0.5\,V and 1\,V, these become equal (see band diagram in Fig.~\ref{Fig1_p1}(c) for $V_g= 0.7$\,V). The situation gets inverted for $V_g>1$\,V.

\begin{center}
	\begin{figure}[t]
	\centering
		\includegraphics[width=0.75\columnwidth]{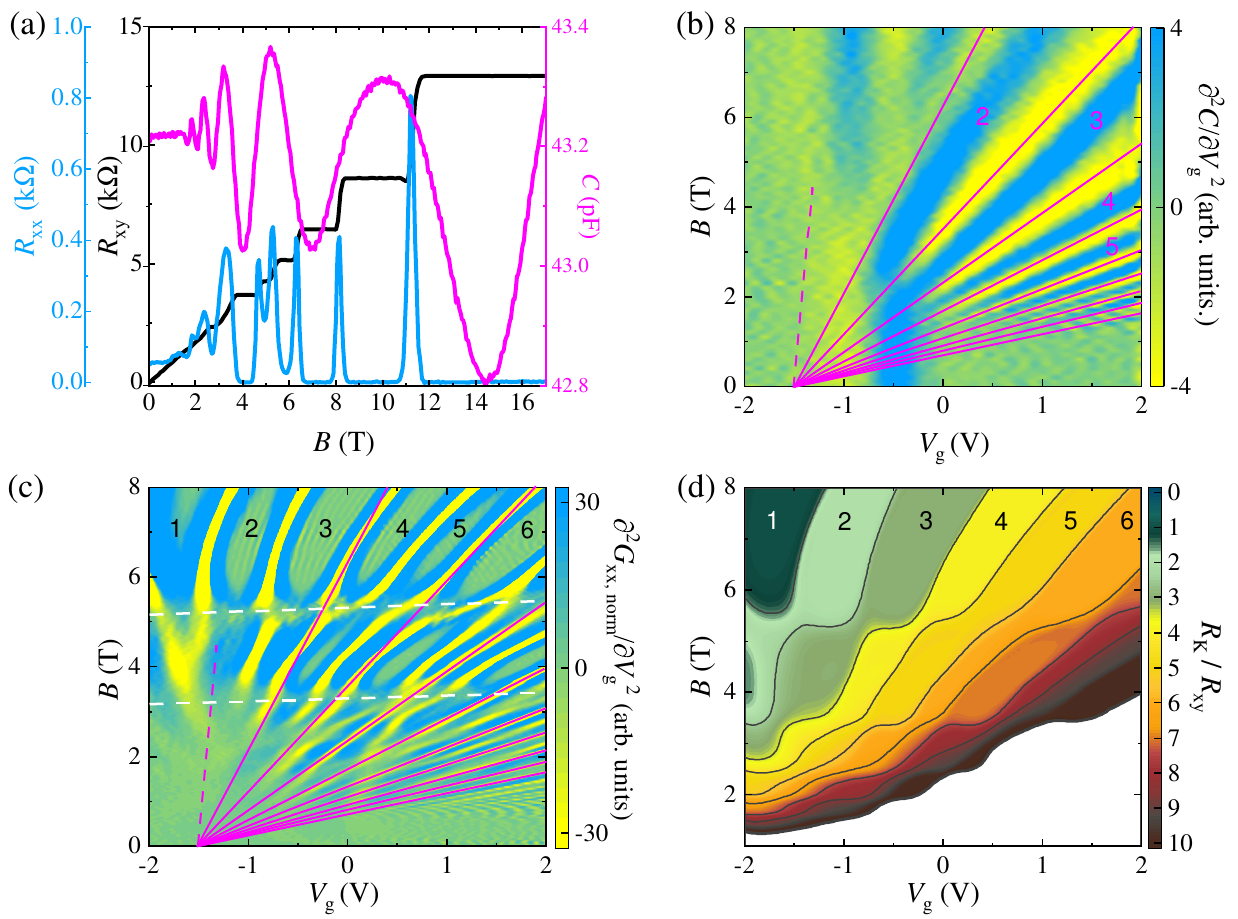}
		\caption{
		(a) Comparison of capacitance (magenta) and transport data for $V_g= 0.5$\,V. Transport data were taken at $T= 57$\,mK and an AC current of 10\,nA, the capacitance was measured at 5\,K and at a AC voltage amplitude of 100\,mV$_{\rm RMS}$ at 3.1\,Hz to minimize resistive effects (see supplemental material to \cite{Kozlov2016}). The minima of $R_{\rm xx}$ match only for some filling factors (2, 4, 7 and higher) with minima in $C$. For other filling factors, minima in capacitance are missing, reflecting different carrier densities being measured by both techniques. 
		(b) Evolution of the capacitance as a function of $V_g$ and $B$, measured at 5\,K, represented in a color map by its second derivative to enhance feature visibility. Yellow corresponds to capacitance maxima, i.e., DOS maxima which are expected for a half-filled Landau level (LL maxima) in the conductive layer closest to the top gate. The magenta lines follow the LL maxima centers. 
		(c) Color map of the second derivative of $G_{\rm xx}$ normalized to its average value with respect to the gate voltage at fixed magnetic field values. Yellow regions correspond to $G_{\rm xx,norm}$ maxima indicating also LL maxima. In greenish regions, $G_{\rm xx}$ (as $R_{\rm xx}$) goes to zero. The blurred, nearly horizontal lines at about 3.3\,T and 5.2\,T, marked by white dashed lines, reflect half-filled LLs on the back surface. The magenta lines are the same as the ones in (b) reflecting only the conductive layer closest to the top gate. At high gate voltages and magnetic fields, these lines correspond to the evolution of certain maxima in $G_{\rm xx}$. 
		(d) Ratio of the Klitzing constant $R_{\rm K}$ to Hall resistance $R_{\rm xy}$ as a function of magnetic field and $V_g$. The different colors reflect the different Hall plateaus. The boundary lines are plotted at the halfway point between two quantum Hall steps. The numbers indicate the relevant integer filling factor. They are also marked in (c). 			
		}
		\label{Fig4}
	\end{figure}
\end{center}

\begin{center}
	\begin{figure}[t]
		\centering
		\includegraphics[width=0.7\columnwidth]{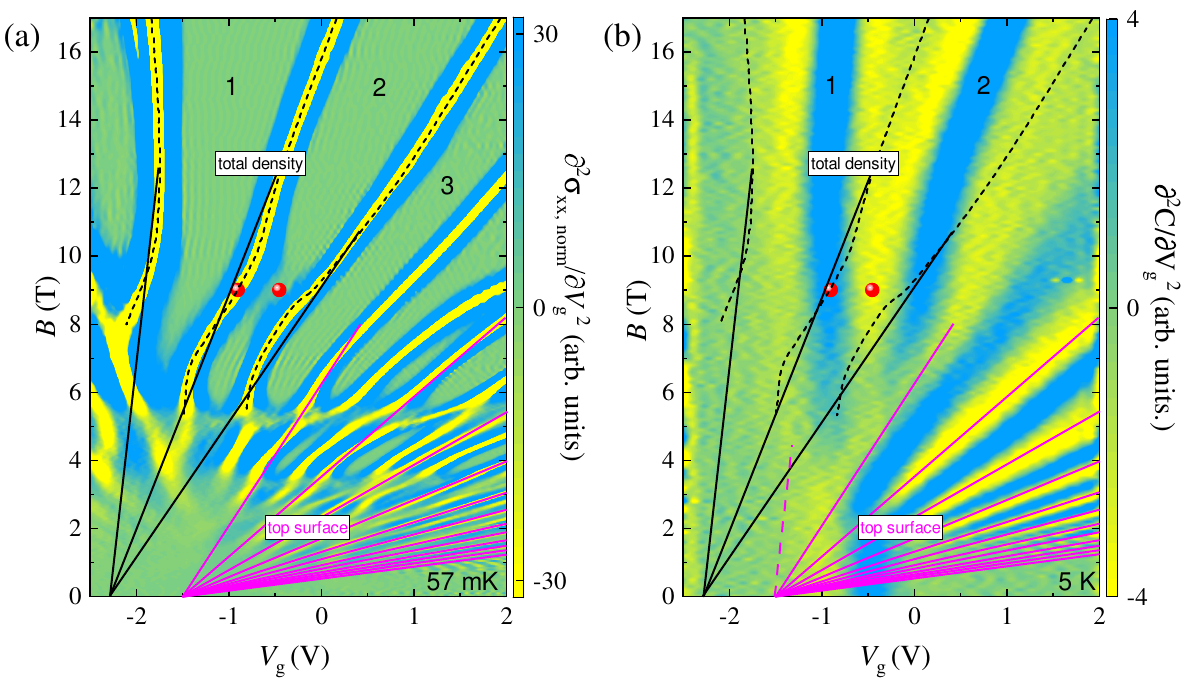}
		\caption{(a) Normalized $\partial ^2 G_{\rm xx,norm}/\partial  V_g^2$.
			color map as in Fig.~\ref{Fig4}(c) but extended to higher $B$. The magenta lines are the same as in Fig.~\ref{Fig4} and are the Landau fan of the top surface electrons, identified by the capacitance measurements at low magnetic field. The dashed lines follow the contour of $\partial ^2 G_{\rm xx,norm}/\partial  V_g^2$, reflecting LL maxima. Above about 12\,T these lines become straight. When the first and the second LL are extrapolated to zero field (solid black lines), they meet at -2.3\,V, i.e. at the CNP. They reflect the Landau fan of the merged electron density. The 0th LL deviates from a straight line, likely due to a nonlinear magnetic field dispersion. The indicated filling factors are those of the merged electron density. (b) Second derivative of the capacitance data with yellow regions showing the density of states maxima. The dashed lines are the ones taken from (a). At high field the LL maxima of capacitance and  $\partial ^2 G_{\rm xx,norm}/\partial  V_g^2$    coincide, indicating that both measurements reflect the same carrier density. The red spheres in both panels indicate the special points where the maximum and minimum of capacitance and conductance are in antiphase (see text).
		}
		\label{Fig5}
	\end{figure}
\end{center}

To emphasize that transport and capacitance measurements give different charge carrier densities due to the existence of different carrier subsystems, spatially separated in the HgTe layer, the corresponding data of $R_{\rm xy}$, $R_{\rm xx}$, and $C$ are directly compared in Fig.~\ref{Fig4}(a). The data are shown for $V_g= 0.5$\,V, where – due to previous analysis – the top layer density and the one at the bottom are close to be equal (at low magnetic fields). 
The capacitance shows pronounced quantum oscillations, but the $1/B$ periodicity of these oscillations starting at low fields is different from that of the SdH oscillations in $R_{\rm xx}$. To examine the transition from low to high magnetic field and from negative to positive $V_g$ in more detail, we compare capacitance data with transport data as a function of magnetic field $B$ and gate voltage $V_g$ in Fig.~\ref{Fig4}(b) and Fig.~\ref{Fig4}(c). 
To enhance the visibility of the maxima in $C$ and $R_{\rm xx}$  at low $B$, we plot the second derivative of the capacitance, $\partial ^2 C/\partial  V_g^2$, and the second derivative of the square magnetoconductance, $\partial ^2 G_{\rm xx,norm}/\partial  V_g^2$.
Here, the $G_{\rm xx}$ data was calculated from $R_{\rm xx}$ and $R_{\rm xy}$ and then normalized to its mean value $G_{\rm xx,norm}$ with respect to the gate voltage at fixed magnetic field values before taking the second derivative. In both graphs, the yellow regions correspond to maxima, in the case of Fig.~\ref{Fig4}(b) capacitance maxima, and for Fig.~\ref{Fig4}(c) $G_{\rm xx}$ maxima. 
Although the quantum oscillations were measured on one and the same device, the resulting Landau fans look strikingly different. The capacitance data show a simple Landau fan in the $(V_g, B)$ plane, the pattern of a 2DES with linearly increasing density in $V_g$. 
This is shown by the magenta lines in Fig.~\ref{Fig4}(b). The lowest Landau level, marked by the dashed magenta line is not resolved in the experiment because of the vicinity to the edge of the valence band at that negative $V_g$ (see band diagram in Fig.~\ref{Fig1_p1}(c)) and therefore the abrupt change of the zero-field $C(V_g)$ in this region. The transport data in Fig.~\ref{Fig4}(c), on the other hand, show a complicated LL pattern. Plotting the Landau fan of Fig.~\ref{Fig4}(b) in Fig.~\ref{Fig4}(c) shows that some of the LL maxima of $G_{\rm xx}$  (as $R_{\rm xx}$) coincide with the magneta LL fan in certain sections. 
These LLs we ascribe to the top surface. The LL sections that do not coincide with the magenta fan must come from the bottom surface and the bulk electrons. 
The blurred narrow regions, lying nearly horizontally in this map just at about $B = 3.3$\,T and $5.2$\,T, highlighted by white dashed lines, can be correlated with quantum oscillations of the conductive bottom surface as the inverse periodicity $\Delta(1/B)$ of these line positions corresponds to the carrier density of the bottom surface, extracted from the Fourier transform shown in Fig.~\ref{Fig3}(b). 
In parallel, the bottom electron system also leaves their mark in the map of the total filling factor, reflected in the inverse Hall resistance. This is shown in Fig.~\ref{Fig4}(d) as $R_{\rm K}/R_{\rm xy}$, with the Klitzing constant $R_{\rm K}$. Close to 3.3\,T and 5.2\,T, steps appear which we also ascribe to the presence of backside electrons. The different quantized Hall plateaus appear in this plot as single-color regions, where the plateau values allow the assignment of the filling factors in Fig.~\ref{Fig4}(c) and Fig.~\ref{Fig5}. The plateau values for filling factors 2 and 3, measured by using standard lock-in techniques, deviate by approximately 0.15\% from the theoretical value, for filling factors 4 to 6 the deviation is between 0.3\% and 0.7\%.

\onecolumngrid
\twocolumngrid

While the Landau fan in Fig.~\ref{Fig4}(c) looks quite complicated with Landau levels running crosswise, the pattern becomes simpler as the magnetic field is increased. This is shown in Fig.~\ref{Fig5} where we again compare capacitance and transport data but now with an extended magnetic field range up to 17\,T. At higher $B$ the Landau fan extracted from the transport data (Fig.~\ref{Fig5}(a)) takes on the look of a conventional 2DES. 
The positions of the LL maxima of the three lowest filling factors are marked with dashed lines. Extrapolation of the dashed lines for the first and second LL maxima to $B = 0$ gives the solid black lines. Their slopes correspond to the filling rate we obtain from the classical $B$-linear Hall voltage, plotted in the Appendix (Fig.~\ref{FigA1}). Thus we associate this filling rate with the one of the total electron density $d n_{\rm tot}/d V_g$. 
This black Landau fan, along with the dashed lines, is copied to Fig.~\ref{Fig5}(b). In contrast to the low-field situation, shown, e.g., in Fig.~\ref{Fig4}, the capacitance maxima now all correspond to the conductivity maxima. 
This means that capacitance and $G_{\rm xx}$ now reflect the same filling rate determined by the same electron density, i.e., the sum of all electron densities. Thus, the central conclusion from these experiments is the following: At low $B$, up to about 6\,T, different Landau fans for the top surface, bottom surface, and conduction band electrons coexist and cause the complicated transport Landau fan. As the magnetic field increases, the Landau fans merge into one, indicating that at sufficiently high magnetic fields only one merged 2D electron system survives in the whole HgTe layer. In this regime, we observe well-quantized Hall plateaus and zero resistance states in $R_{\rm xx}$, as shown in Fig.~\ref{Fig2}(a).

\onecolumngrid
\twocolumngrid

In the simplest terms, the merging of the top and bottom surface electrons with the bulk electrons into a single two-dimensional electron system (2DES) can be understood by examining the distinct evolution of Landau levels for these three subsystems. At zero magnetic field, the different electron densities in each subsystem result in different initial Fermi energies. However, the overall Fermi level remains the same across the three subsystems. As the magnetic field increases, the LL degeneracy increases linearly with the magnetic field, and therefore the energetically higher LL levels in each subsystem get depopulated. Energetically driven, electrons move into the lower LL of the subsystems or change to other subsystems as all are electrically connected. In the latter case, the overall electrostatics is affected. The exchange of electrons between the subsystems might lead to an energetically more favorable situation. To model this, one must consider self-consistently the modification of the electrostatics, as well as the dispersion of the different Landau levels, the effect of Landau level crossings, and the spatial distribution of the bulk electrons. The lack of knowledge of some of these parameters makes the modeling unreliable. 
At a sufficiently high magnetic field, only the evolution of a single LL fan is visible in the $B-V_g$ diagrams of Fig.~\ref{Fig5}. By linear extrapolation, the corresponding LL fan has its origin at ($B= 0$\,T, $V_g= -2.3$\,V). 
At high magnetic fields, this merged 2DES acts as the counter electrode to the gate electrode. The fact that the electron density extracted from the capacitance at high magnetic fields corresponds to the total carrier density and can be directly tuned by the gate voltage supports the assumption that only a single 2DES remains. The merging of a top and bottom electronic subsystem with an increasing magnetic field has also been observed in wide quantum well structures based on (Al,Ga)As heterostructures \cite{Dorozhkin2023}.

Finally, we looked at the non-equilibrium current distribution in the quantum Hall regime. Similar to investigations over the last two decades on quantum Hall samples based on (Al,Ga)As heterostructures \cite{Weitz2000, Weitz2000-2, Ahlswede2001, Ahlswede2002, Dahlem2010, Panos2014} and exfoliated graphene \cite{Panos2014PhD}, this was obtained by measur-ing the local Hall voltage drop over the cross-section of the Hall bar by a scanning probe microscope sensitive to electrostatic potential variations within the 2DES. 
The scheme of the experimental setup is presented in Fig.~\ref{Fig1_p2}(c). The measurements were conducted in a top-loading $^3$He-$^4$He dilution fridge at a base temperature of 40\,mK on an ungated Hall bar with a width of 20\,$\mu$m, fabricated from the same heterostructure as previously described. A field effect transistor – fabricated from an (Al,Ga)As heterostructure containing a 2DES – was used here as a local probing sensor. 
Due to the orientation of the sensor relative to the Hall bar, at best a spatial resolution of 1\,$\mu$m in $x-$ and 10\,nm in $y$-direction can be expected. The Hall potential profile normalized to the applied Hall bar bias voltage is obtained by scanning twice at a distance of about 0.3\,$\mu$m over the cross-section of the Hall bar with different voltage excitations applied to the Hall bar source and drain contact (see in Fig.~\ref{Fig1_p2}(c) the switch position a or b) \cite{McCormick1999,Weitz2000}.

\begin{figure}[t]
	\centering
	\includegraphics[width=1\columnwidth]{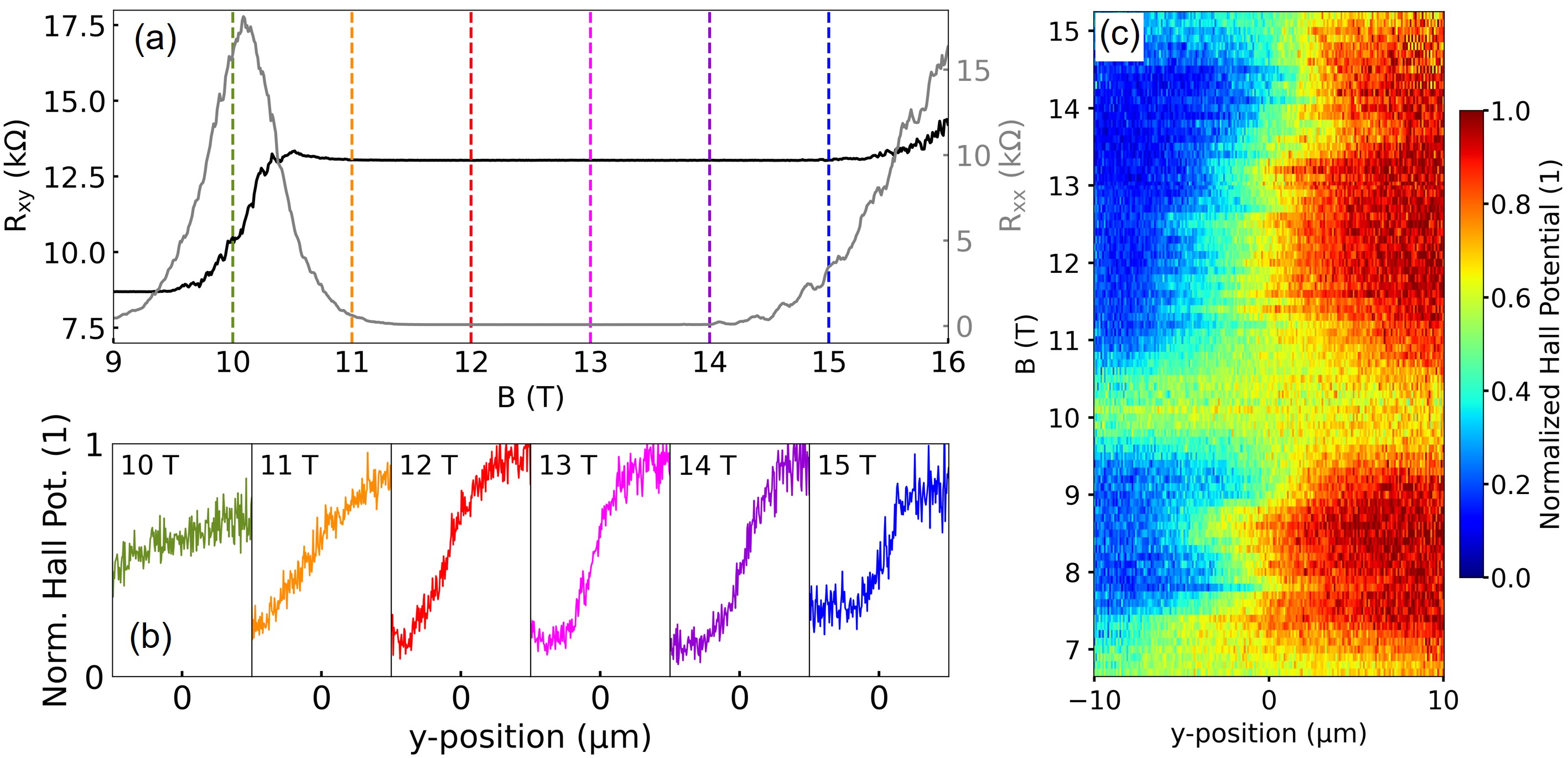}
	\caption{(a) Magneto-transport data of the ungated Hall bar, measured at 40\,mK and a bias voltage of $V_{\rm DS} = 0.3$\,mV. The dotted lines indicate the magnetic field positions in the plateau for which the Hall potential profiles are shown in (c). (b) Hall potential profiles with-in the $\nu = 2$ QH plateau, normalized to the applied voltage, measured over the width of the Hall bar in the middle between potential probing contacts. (c) Color map of Hall potential profiles for a larger magnetic field range containing the plateaus of $\nu = 2$ (in the range of $B = 11\ldots15$\,T) and $\nu = 3$ ($8\ldots9$\,T).
	}
	\label{Fig6}
\end{figure}

Fig.~\ref{Fig6}(a) shows $R_{\rm xx}$ and $R_{\rm xy}$ vs. $B$ around the Hall resistance plateau with $\nu = 2$ measured on the ungated sample with $V_{\rm DS} = 0.3$\,mV. The magnetic field range where this plateau appears fits very well to the magneto-transport characteristics obtained on the gated Hall bar at $V_g = 0$. Please note, the Hall resistance plateau is more extended than the $R_{\rm xx}$ zero-resistance state which is measured along the Hall bar with 150\,$\mu$m distance between the potential probing contacts. This indicates that here the charge carrier density is slightly inhomogeneous along the Hall bar.

In Fig.~\ref{Fig6}(b), Hall potential profiles, normalized to the applied bias voltage of 0.3\,mV, are shown, measured over the cross-section at about half the distance between the potential probing contacts for different magnetic field values in the quantum Hall plateau. For $B = 10$\,T, i.e., in the transition regime between quantum Hall plateaus, the Hall potential profile shows a linear Hall potential drop over the width of the Hall bar, as expected for a classical Hall effect. 
The Hall potential difference between left and right side is much less than the applied bias voltage (normalized Hall voltage is about 20\%) which is mainly due to the voltage drop along the resistive Hall bar. 
For $B = 11$\,T, the sample enters the plateau from the lower magnetic field side. The Hall resistance is well quantized, yet the longitudinal resistance is still finite. Looking at the respective Hall potential profile in Fig.~\ref{Fig6}(b), the potential drop shifts to the center of the cross-section, increasing in steepness and overall magnitude (to about 70\% of the applied voltage). Increasing the magnetic field to $B = 12$\,T, $R_{\rm xx} = 0$ is reached. The Hall potential profile for $B =12$\,T, 13\,T, and 14\,T shows a Hall voltage of 90\% of the applied bias voltage (the wiring to the QH sample and the resistances at source and drain contact are responsible for the remaining 10\%), and slight variations in the Hall potential drop at the cross-section center. At $B = 15$\,T, the longitudinal resistance shows again a finite value, going hand in hand with a decrease in the Hall voltage seen in this cross-section.

The changes of the Hall potential profiles throughout a plateau can be better seen in Fig.~\ref{Fig6}c. Here, the Hall potential profiles for a larger magnetic field range are shown as a color map, containing the plateaus for filling factors $\nu = 2$  and $\nu = 3$. 
The positions of the potential drop are visible as green-yellow-colored areas.  For both plateaus, a continuous evolution occurs from a linear potential drop over the entire Hall bar width to a potential drop only in the center of the cross-section and back again to a linear drop over the whole cross-section. Inhomogeneities of the electron density along the Hall bar become visible by measuring the Hall potential profile for fixed magnetic field ($B = 14$\,T) at different cross-sections along the Hall bar (see Fig.~\ref{FigA2} in the Appendix): For that magnetic field value within the QH plateau, the Hall potential drop in the center continuously varies from cross-section to cross-section shifting forth and back in the cross-section on a scale of few micrometers along the Hall bar.

This behavior is expected from scanning probe experiments on quantum Hall samples based on (Al,Ga)As heterostructures and their interpretation \cite{Weis2011}. The quantized Hall resistance appears due to Hall voltage drops only over locally incompressible regions, all of the same integer-valued filling factor $\nu=i$. The local drop over an incompressible region in the $y$-direction causes a dissipationless Hall current in the $x$-direction, carried locally by all occupied Landau levels within the incompressible region. This is expressed by a local relation between Hall field $E_y$ and local current density $j_x=i(e^2/h) E_y$.

In contrast to (Al,Ga)As QH samples, here in the HgTe sample, only the bulk-dominated quantum Hall regime is observed. The edge-dominated QH regime, found in (Al,Ga)As QH samples at the lower magnetic field side of a QH plateau, is caused by the pronounced depletion region at the Hall bar mesa edges where the electron density goes down to zero over a typical length of a $\mu$m. Due to that, with increasing magnetic field pronounced incompressible strips appear in the depletion regions at the Hall bar edges before the bulk becomes incompressible. Wide enough, Hall voltage drops over such incompressible strips cause a dissipationless non-equilibrium current flow along both edges of the Hall bar. 
With rising $B$, the innermost incompressible strip at each edge shifts strongest in its positions towards the Hall bar center while getting also wider, until innermost incompressible strips of opposite edges merge to an in-compressible bulk, where – due to inhomogeneities in the charge carrier concentration – compressible droplets are embedded. Hall potential profiles, measured in (Al,Ga)As QH samples, reflect this evolution of incompressible regions from the edges to the bulk, i.e., within a quantum Hall plateau we moved from an edge- to a bulk-dominated QH regime \cite{Weis2011}. This can even be seen in electrically induced QH breakdown measurements \cite{Haremski2020} on (Al,Ga)As QH samples of different Hall bar widths. The topological surface states of HgTe at the mesa edges do not allow for such depletion at the edges of the Hall bar. Therefore, the edge-dominated QH regime is here suppressed, only the robust bulk-dominated QH regime of a merged 2DES exists where the dissipationless current is driven by the local Hall voltage drop in incompressible regions of the 2DES. In the QH regime, the compressible edges have the important role of carrying the electrochemical potential difference between source and drain contact into the QH sample which is then present as Hall voltage in the Hall bar cross-section. The topological surface states of HgTe at the mesa edges fulfill this task.

As previously discussed, quantum Hall effect is not only observed at high magnetic field but also in the regime where more charge carrier subsystems co-exist. As seen in Fig.~\ref{Fig4}(d), the quantum Hall plateaus reflect an integer filling factor for the total electron density although the subsystems have not merged to a single one. This is possible if the two Dirac electron systems – the one at the top and the one at the bottom of the HgTe layer – are both incompressible (with compressible droplets embedded) ($G_{\rm xx} = 0$). Due to the electrically conductive side facets, both subsystems are connected and see the same Hall voltage $V_y$. If within a Hall bar cross-section the Hall voltage drops within each subsystem only over incompressible regions of the same filling factor – which can be different for both subsystems, i.e., $\nu_t=i_t$ and $\nu_b=i_b$, the resulting non-equilibrium current is carried dissipationless by both subsystems: $I_x=(i_t+i_b )(e^2/h) V_y$.

In the transition between plateaus we see that maxima of $G_{\rm xx}$ and $C$ do not always coincide. 
For instance, the coincidence of the maximum of $G_{\rm xx}$ and a minimum of $C$ (e.g., at $B=9$\,T and $V_g= -0.9$\,V marked as one of the red spheres in Fig.~\ref{Fig5}) is explained by the fact that under these conditions the electron subsystem on the top surface has a low density, behaves incompressible and therefore only partially screens the gate electrode. Consequently, the deeper-lying compressible subsystem is charged, giving - due to the larger distance to the gate - less contribution to the capacitance. The coexistence of zero $G_{\rm xx}$ and a capacitance maximum (e.g., at $B=9$\,T and $V_g= -0.45$\,V) suggests a different situation. Zero $G_{\rm xx}$ hints to the fact that all subsystems are incompressible. While most of the additional charge carriers are induced in the top layer at that gate voltage, obviously these electrons remain localized in compressible droplets. The various situations in the data can be interpreted – more or less stringent - in that way, however a detailed model to describe the respective evolution of $G_{\rm xx}$ and $C$ is not yet available. 

In conclusion, we have studied the quantum Hall effect in a 3D topological insulator, specifically a strained HgTe film, by combining different measurement techniques. At low magnetic fields, multiple electron subsystems are identified, that develop distinct Landau fans. Quantum Hall plateaus, corresponding to integer-valued Landau level filling factors of the total electron density, are consistently observed in this regime. Our results at high magnetic fields indicate that these subsystems merge into a single electron system at sufficiently high magnetic fields. In this regime, the non-equilibrium current distribution has been probed using spatially resolved measurements of the Hall potential with an electric field-sensitive scanning probe technique. These measurements show that the side facets of the conductive mesa conduct the bias voltage applied between the source and drain contacts along the Hall bar edges into the sample. In the quantum Hall regime, this bias voltage is present as Hall voltage in the cross-section of the Hall bar and drives a dissipationless current through the incompressible bulk – a quantized Hall resistance is measured. This suggests what happens at lower magnetic fields, where there are still electron subsystems at the top and bottom of the HgTe layer: both subsystems see the same Hall voltage and - if both subsystems are incompressible in the bulk - each carry a fraction of the dissipationless current.

\bibliography{references}

\subsection{Appendix}


\setcounter{figure}{0}
\renewcommand{\thefigure}{A\arabic{figure}}

\begin{figure}[h]
	\centering
	\includegraphics[width=1\columnwidth]{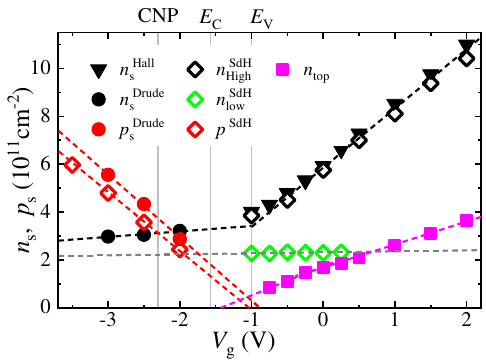}
	\caption{Charge carrier densities of the different charge carrier types. Black data points are the total electron density $n_{\rm tot}$ at particular gate voltage, extracted from the linear Hall slope ($n_s^{\rm Hall}$), from high-field SdH-oscillations ($n_{\rm high}^{\rm SdH}$) and, on the hole side, from the two-carrier Drude model ($n_s^{\rm Drude}$). The valence band hole densities, shown in red were extracted using the two-carrier Drude model ($p_s^{\rm Drude}$) and SdH-oscillations ($p^{\rm SdH}$). Top surface electron density ($n_{\rm top}$) is obtained from the period of the capacitance oscillations and the back surface electron density $n_{\rm bot}$ by Fourier analysis of the low field SdH-oscillations.
	}
	\label{FigA1}
\end{figure}

Fig.~\ref{FigA1} summarizes the densities of the different charge carrier species as a function of $V_g$, extracted by different techniques. Similar data for other devices has been discussed previously \cite{Kozlov2014,Ziegler2020,Kozlov2016}. The carrier density measured by the classical, $B$-linear Hall voltage for $V_g> -1$\,V, $n_s^{\rm Hall}$, involves all electrons. Therefore we call it the total electron density $n_{\rm tot}=n_s^{\rm Hall}$. The carrier density extracted from periodicity of the SdH oscillations measured at high magnetic fields, $n_{\rm high}^{\rm SdH}$, coincides with $n_s^{\rm Hall}$ for $V_g> -1$\,V, so that $n_{\rm high}^{\rm SdH} = n_{\rm tot}$ holds. In the valence band ($V_g< -1$\,V), where electrons and holes coexist, the sum of top and bottom surface electrons, $n_s^{\rm Drude}$, is extracted from the two-carrier Drude model. The total hole density $p_s^{\rm Drude}$ is also obtained from the two-carrier Drude model. SdH measurements on the valence band show a slightly lower hole density $p^{\rm SdH}$ than the Drude result, a feature, which is explained by the partial compensation of electrons and holes \cite{Kozlov2014,Mendez1985}. The capacitance quantum oscillations, as discussed in the main text above, reflect for not too high $B$ only the carrier density $n_{\rm top}$ of the Dirac top surface electrons; the $V_g$ dependence of the respective carrier density is shown by the magenta symbols in Fig.~\ref{Fig3}. In contrast, the green data points show $n_{\rm bot}(V_g)$. These values were extracted from the data in Fig.~\ref{Fig3}.

\begin{figure}[h]
	\centering
	\includegraphics[width=1\columnwidth]{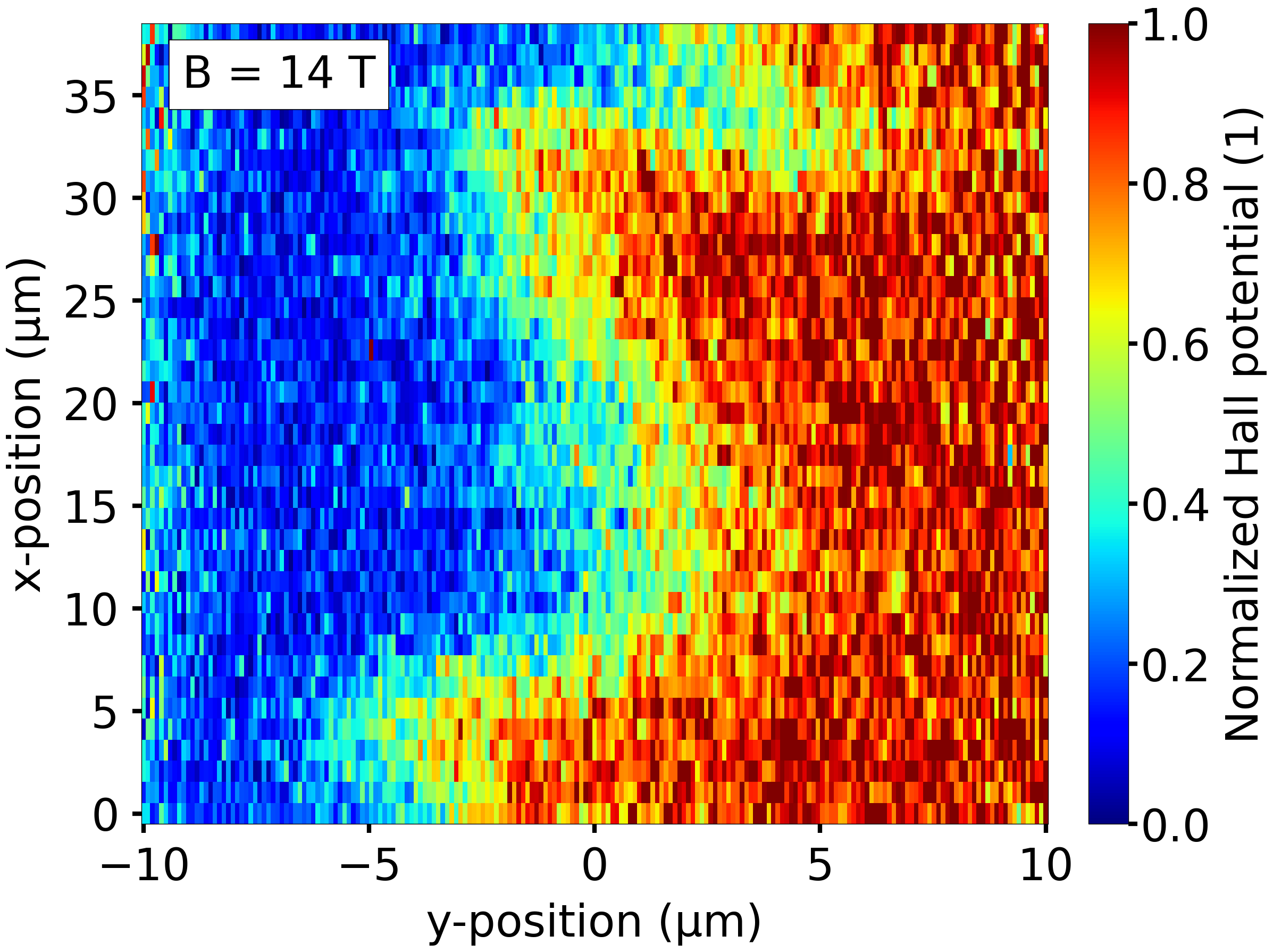}
	\caption{Hall potential distribution over an area of the Hallbar for a magnetic field of $B = 14$\,T, being in the well quantized regime of QH plateau $\nu=2$.
	}
	\label{FigA2}
\end{figure}

\end{document}